\documentclass{PoS}

\usepackage{amsmath}
\usepackage{amssymb}
\usepackage{amsfonts}
\usepackage{graphics}
\usepackage{graphicx}
\usepackage{amscd}
\usepackage{amsfonts}
\usepackage{epsf}
\usepackage{epsfig}
\usepackage{color}
\usepackage{bm}

\def\be#1{\begin{equation}#1\end{equation}} %{equation}
\def\beqnn#1{\begin{eqnarray}#1\end{eqnarray}}
\def\exp#1{\mathrm{e}^{#1}} %???R????
\def\integral#1#2{\int^{#2}_{#1}} %?C???e?O????
\title{Charmonium contribution to $B\to K\ell^+\ell^-$: testing the factorization approximation on the lattice}

\ShortTitle{$B\to K\ell^+\ell^-$ Factorization}

\author{\speaker{Katsumasa Nakayama}$^{a, b}$, Tsutomu Ishikawa$^{b,c}$, and Shoji Hashimoto$^{b,c}$ (JLQCD collaboration)\\
   $^a$ NIC, DESY Zeuthen, Platanenallee 6, 15738 Zeuthen, Germany\\
   $^b$ KEK Theory Center, High 	Energy Accelerator Research Organization (KEK), Tsukuba 305-0801, Japan\\
   $^c$ School of High Energy Accelerator Science, The Graduate University for Advanced Studies (Sokendai),Tsukuba 305-0801, Japan\\
  E-mail: \email{katsumasa.nakayama@desy.de}}

\abstract{
We report the current status of a study of charmonium contribution to $B \rightarrow K\ell^+\ell^-$ on the lattice.
Our lattice calculation tests the factorization approximation for this contribution.
In order to control the problem of the artificial divergence, we focus on the low $q^2$ region with a small b-quark mass.
We also take into account the renormalization constants of relevant four-quark operators calculated through the temporal moments.
Results suggest a violation of the factorization approximation.
}

\FullConference{The 37th Annual International Symposium on Lattice Field Theory - LATTICE2019\\
  16-22 June, 2019\\
  Wuhan, China.}

\begin{document}

\section{Introduction}

The rare decay $B \rightarrow K^{(*)}\ell^+\ell^-$ has received much attention as a clean probe of new physics since the Standard Model contribution is suppressed due to flavor-changing neutral-current.
Sizable difference from the Standard Model has been reported for the differential decay rate of $B \rightarrow K^{(*)}\ell^+\ell^-$ by LHCb \cite{Aaij:2013pta, Aaij:2016cbx}.

In order to confirm this tension, we have to control the uncertainty due to non-perturbative contributions.
The experimental analysis of the $B\to K^{(*)}\ell^+\ell^-$ decays focused on the region where invariant mass squared of the finial lepton pair $q^2$ is not close to the charmonium resonances.
However, long-distance effects between the final state kaon and the virtual charmonium state could be significant even outside such resonance regions.

So far, theoretical estimates have been attempted by using the perturbative calculation and applying the factorization approximation, although the intermediate state can be more complex.
In the factorization, we approximate the amplitude by a product of the $B\to K$ part and the charmonium resonance part.
In other words, we ignore the interaction between the $B\to K$ form factor and the charmonium two-point function.
The factorization approximation has been studied with experimental results and models, but reliable prediction for the $B\to K$ decay remains to be difficult \cite{Neubert:1997uc,Beneke:2001at,Lyon:2014hpa,Du:2015tda}.

In this proceedings, we report the recent progress of the numerical lattice calculation to study the factorization approximation for the $B \rightarrow K\ell^+\ell^-$ amplitude. 
We calculate the $B \rightarrow K\ell^+\ell^-$ decay amplitude with and without the factorization.
We take account of the renormalization constant and provide a test of the factorization approximation using an explicit lattice calculation.

\section{$B\to K \ell^+\ell^-$ amplitude and the artificial divergence}
In this section, we review the calculation of the decay amplitude with special emphasis on the artificial divergence.
Avoiding such divergence is essential for the lattice calculation, and the problem is extensively studied for the calculation of $K\to \pi \ell^+\ell^-$ amplitude on the lattice \cite{Christ:2015aha,Christ:2016eae}.

We consider the $B \to K$ amplitude with the charmonium contribution, which occurs through the weak effective Hamiltonian $H_\mathrm{eff}$ with the Fermi constant $G_F$, CKM matrix $V_{cb},V_{cs}$, and Wilson coefficient $C_i$,
\be{
H_\mathrm{eff}
=
\frac{G_F}{\sqrt{2}}
V_{cs} ^*V_{cb}
\left(
C_1O_1 ^c
+
C_2O_2 ^c
\right).
}
The operators $O_i ^c$, which include $c\overline{c}$ are
\beqnn{
O_1 ^c
&=&
(\overline{s}_i\gamma_\mu P_-c_j)
(\overline{c}_j\gamma_\mu P_-b_i),\nonumber\\
O_2 ^c
&=&
(\overline{s}_i\gamma_\mu P_-c_i)
(\overline{c}_j\gamma_\mu P_-b_j).
}
where indeces $i$ and $j$ represent the color index, and the chiral projection operator is defined as $P_- \equiv \frac{1 - \gamma_5}{2}$.

We define the $B\to K\ell^+\ell^-$ decay amplitude for a four-momentum $q \equiv k - p$ as
\be{
A(q^2)=\int\mathrm{d}^4x\ 
\exp{iqx}
\langle
K(\bm{p})|
T\left[
J_\mu(0)
H_\mathrm{eff}(x)
\right]
|B(\bm{k})
\rangle.
}

In order to calculate the amplitude, we integrate over the position of the weak effective Hamiltonian and define $I_\mu$ as
\be{
I_\mu
%(T_a,T_b,\bm{k},\bm{p})
=
\exp{-\left[E_K(\bm{p}) - E_B(\bm{k})\right]t_J}
\integral{t_J - T_a}{t_J + T_b}
\mathrm{d}t_H
\integral{}{}\mathrm{d}^3\bm{x}
\integral{}{}\mathrm{d}^3\bm{y}
\ \exp{-i\bm{q}\cdot\bm{x}}
\langle
K(t_K,\bm{p})
|T\left[
J_\mu(t_J,\bm{x})
H_\mathrm{eff}(t_H,\bm{y})
\right]|B(0,\bm{k})
\rangle.
}
\begin{figure}[tbp]
\begin{center}
 \includegraphics[width=6cm, angle=0]{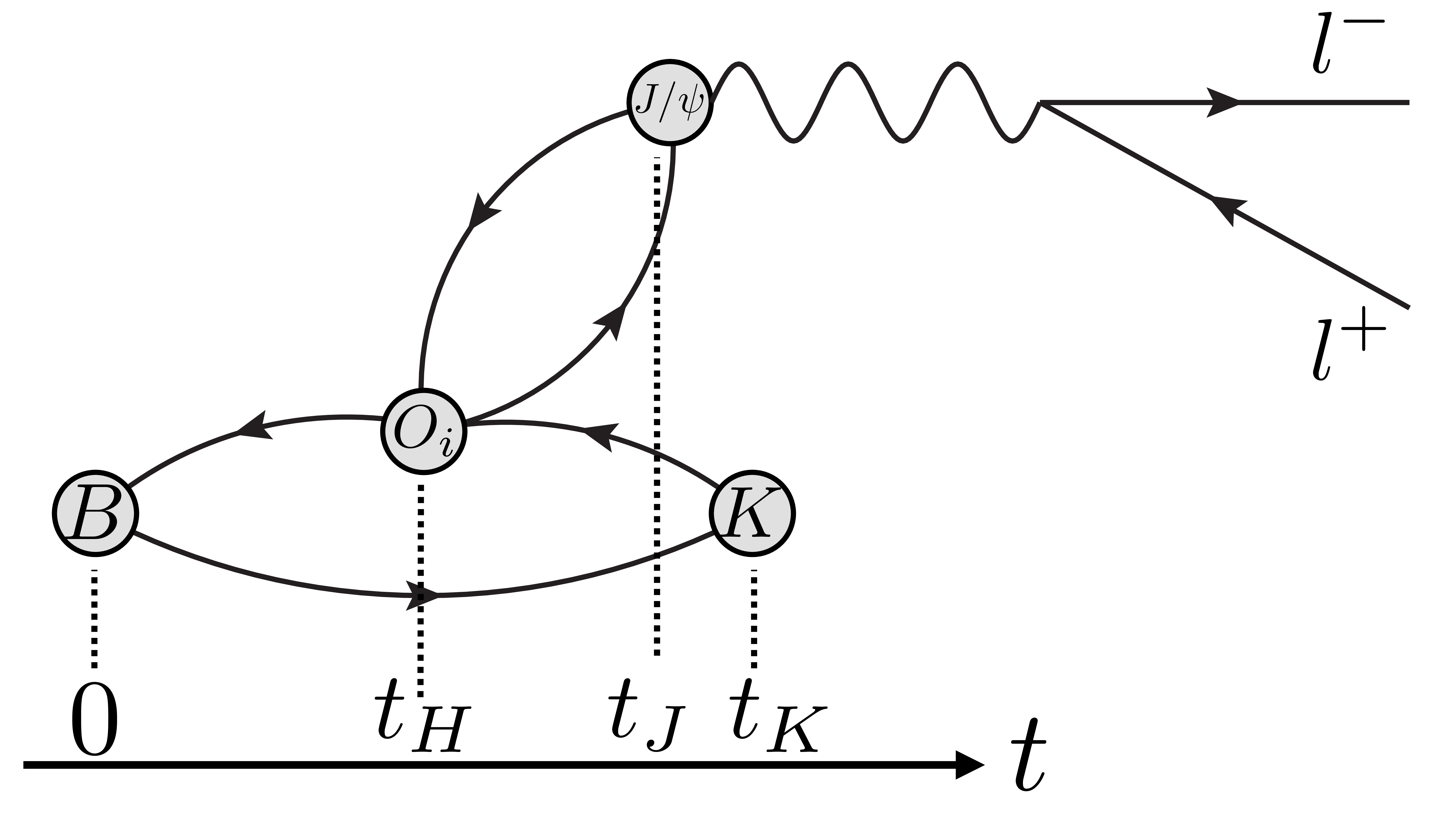}
 \caption{
 Setup of the lattice calculation of the $B\to K \ell^+\ell^-$ amplitude through charmonium $J/\psi$ resonances.
}
\label{fig:BtoKll}
\end{center}
\end{figure}
The setup of the lattice calculation is shown in Figure \ref{fig:BtoKll}.
We introduce $t_H, t_J, t_K, T_a,$ and $T_b$ to identify the time for each states and operators.

We can rewrite this quantity using the complete set of the intermediate states, which can be described by the spectral densities $\rho_1(E)$ for the states with strangeness, and $\rho_2(E)$ for those without strangeness.
Namely, 
\beqnn{
I_\mu
%(T_a,T_b,\bm{k},\bm{p})
&=&
-\integral{0}{\infty}\mathrm{d}E
\frac{\rho_1(E)}{2E}
\frac{
\langle K(\bm{p})|J_\mu(0)|E(\bm{k})\rangle
\langle E(\bm{k})|H_\mathrm{eff}(0)|B(\bm{k})\rangle
}
{
E_B(\bm{k})-E
}\left(
1 - \exp{(E_B(\bm{k})-E)T_a}
\right)
\nonumber\\
&&+
\integral{0}{\infty}\mathrm{d}E
\frac{\rho_2(E)}{2E}
\frac{
\langle K(\bm{p})|H_\mathrm{eff}(0)|E(\bm{p})\rangle
\langle E(\bm{p})|J_\mu(0)|B(\bm{k})\rangle
}
{
E-E_K(\bm{p})
}\left(
1 - \exp{-(E-E_K(\bm{p}))T_b}
\right).
\label{int_div}
}
In this representation, the $T_{a,b} \to \infty$ limit of $I_\mu$ can be identified as the amplitude,
\be{
A(q^2)
=
-i\lim_{T_{a,b}\to \infty}
I_\mu(T_a,T_b,\bm{k},\bm{p}).
}

In order that the integral (\ref{int_div}) stays finite, the energy of the intermediate state plays an essential role.
Since $E - E_K(\bm{p}) > 0 $ is always satisfied, $\exp{-(E - E_K(\bm{p}))T_b}$ can be ignored in the $T_b\to \infty$ limit.
On the other hand, $E_B(\bm{k}) - E < 0$ is not always satisfied, depending on the intermediate energy and the term $\exp{(E_B(\bm{k}) - E)T_a}$ may diverge in the limit of large $T_a$. 
At the physical point of the quark masses, this artificial divergence can be hardly removed, since the number of such intermediate states is large.
In this study, we set the b-quark mass smaller than that of the physical value in order to avoid this problem.
Since the energy of the intermediate state $E$ is bounded by the ground state energy of the K and $J/\psi$ meson, we choose the b-quark mass to realize the condition, $E_B < E_{J/\psi} + E_K$.

With this unphysical b-quark mass, we can define the decay amplitude from the four-point correlators.
In this work, however, we test the factorization approximation as the first step before proceeding to the extraction of the decay amplitude.

\section{Factorization and renormalization}

\begin{figure}[tbp]
\begin{center}
 \includegraphics[width=8cm, angle=0]{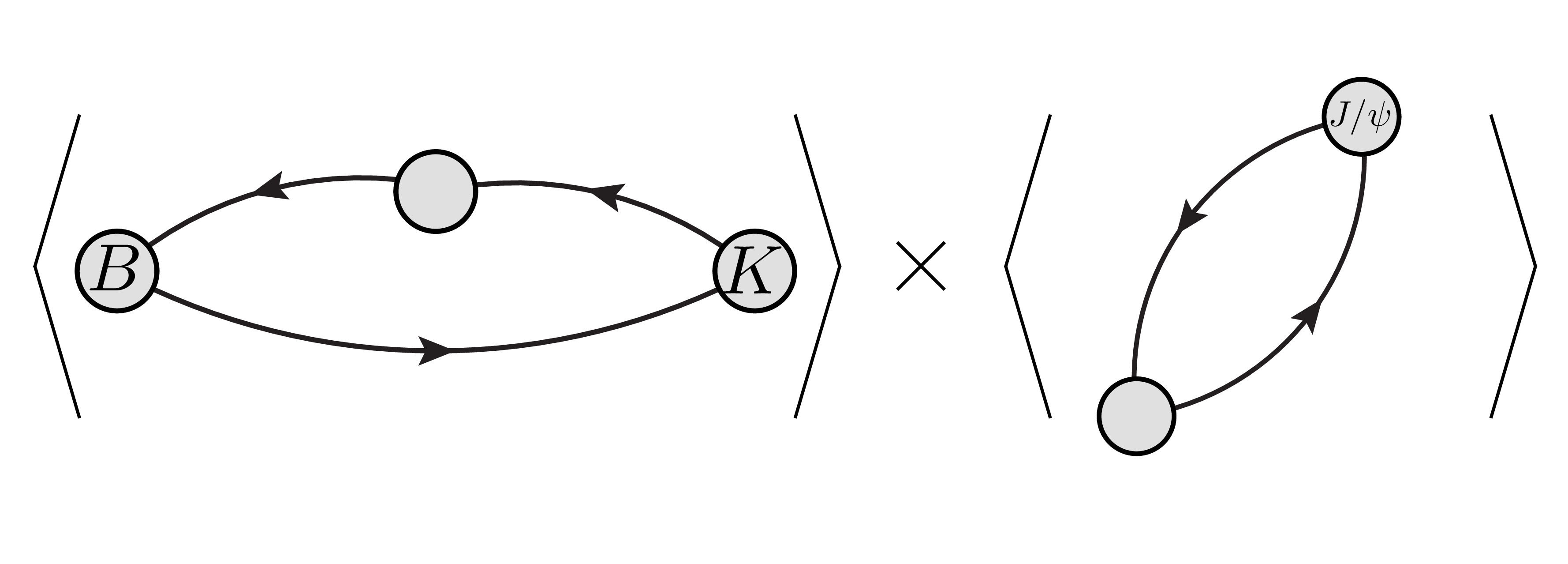}
 \caption{
 Factorization of the four-point correlator $B\to K \ell^+\ell^-$ with charmonium $J/\psi$ resonances.
}
\label{fig:factori}
\end{center}
\end{figure}

\begin{figure}[tbp]
\begin{center}
 \includegraphics[width=4.5cm, angle=0]{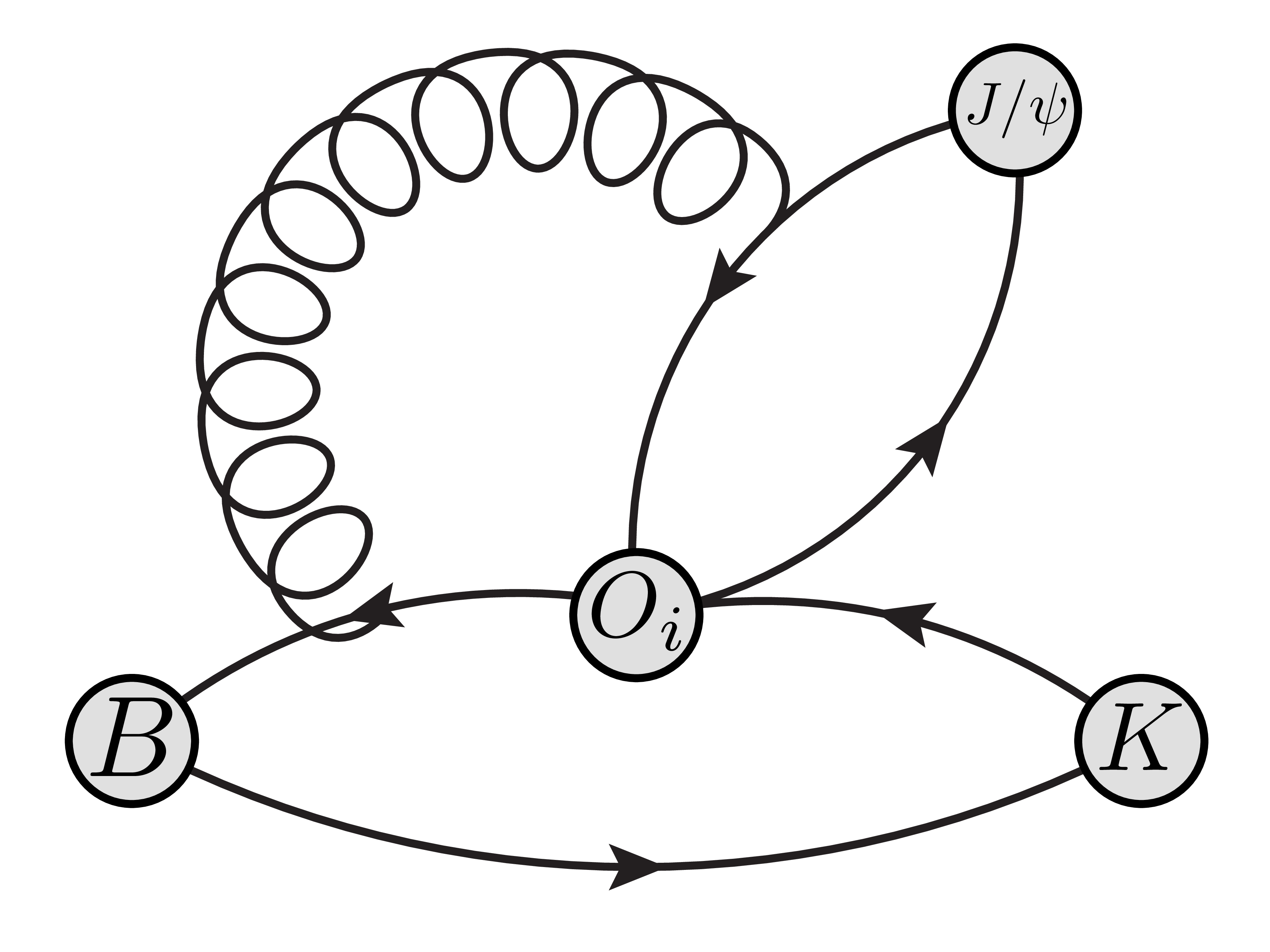}
 \caption{
 A typical example of the non-factorizable contribution for $B\to K \ell^+\ell^-$ with the charmonium.
 Gluon exchanging between $B$ (K) and the charmoinum can not be factorized.
}
\label{fig:onnceg}
\end{center}
\end{figure}

In order to investigate the factorization, we define operators $O^{(1)}$ and $O^{(8)}$ as
\beqnn{
O^{(1)}
&=&
(\overline{c}_i\gamma_\mu P_- c_i )
(\overline{s}_j\gamma_\mu P_- b_j),\nonumber \\
O^{(8)}
&=&
(\overline{c}_i[T^a]_{ij}\gamma_\mu P_- c_j )
(\overline{s}_k[T^a]_{kl}\gamma_\mu P_- b_l).
}
The operator with the color octet contraction $O^{(8)}$ includes the $SU(3)$ generators $T^a$.

Figure \ref{fig:factori} illustrates the factorization of the $B\to K \ell^+\ell^-$ four-point correlator.
The contribution of $O^{(1)}$ is simply represented in the factorization approximation as,
\be{
\langle
KJ/\psi |
O^{(1)}
|B\rangle
\simeq
\langle K |
\overline{s}_i\gamma_\mu P_-b_i
|B
\rangle
\langle J/\psi |
\overline{c}\gamma_\mu c
| 0\rangle.
\label{fac1}
}

Figure \ref{fig:onnceg} is a typical example of the non-factorizable contribution.
As we can see in the definition of $O^{(8)}$, the simple factorization is not allowed for this operator, because the factorized piece is color-octet, which vanishes when sandwiched by the physical states.
Namely, factorization of the $O^{(8)}$ is represented as 
\be{
\langle
KJ/\psi |
O^{(8)}
|B\rangle
\simeq
0.
\label{fac2}
}

Since our lattice calculation is done in the $O_1 ^c$ and $O_2 ^c$ basis, we need to transform them to the $O^{(1)}$ and $O^{(8)}$ basis.
The Firtz transformation
$
\overline{q}_1\gamma_\mu P_- q_2 \overline{q}_3\gamma_\mu P_-q_4
=
\overline{q}_1\gamma_\mu P_- q_4 \overline{q}_3\gamma_\mu P_-q_2
$
can be used to obtain the relation between $O_1 ^c, O_2 ^c$ and $O^{(1)},O^{(8)}$:
\beqnn{
O_1 ^c
&=&
O^{(1)},\nonumber\\
O_2 ^c
&=&
\frac{1}{3}
O^{(1)}
+
2O^{(8)}.
}

Here, we also consider the renormalization of the operator $O_1 ^c$ and $O_2 ^c$.
The renormalized operators $\langle O_1 ^c\rangle_R$ and $\langle O_2 ^c\rangle_R$ are written in terms of $O_1 ^c$ and $O_2 ^c$ with the renormalization constants $Z_{11}$ and $Z_{12}$,
\beqnn{
\langle
O_1 ^c
\rangle_R
&\equiv&
Z_{11}
\langle
O_1 ^c
\rangle
+
Z_{12}
\langle
O_2 ^c
\rangle
\nonumber\\
\langle
O_2 ^c
\rangle_R
&\equiv&
Z_{12}
\langle
O_1 ^c
\rangle
+
Z_{11}
\langle
O_2 ^c
\rangle.
}
The renormalization constant are determined through temporal moments of three-point correlators \cite{Ishikawa_proc}.

In order to test the factorization relation, we define the ratios $R_1$ and $R_{1/3}$ on the lattice of volume $V$,
\beqnn{
R_1
&\equiv&
\frac{V\langle K|J_\nu O_1 ^c |B\rangle_R}
{
\langle 0|
J_\nu J_\mu
|0\rangle_R
\langle K |
 \overline{s}_j\gamma_\mu P_- b_j
|B\rangle_R
},
\nonumber\\
R_{1/3}
&\equiv&
\frac{\langle K|J_\nu O_2 ^c |B\rangle_R}
{
\langle K|J_\nu O_1 ^c |B\rangle_R
},
}
which become $1$ or $\frac{1}{3}$ when the factorization approximation is valid, respectively.

\section{Preliminaly results}
\begin{table}
\centering
 \begin{tabular}{cccc|ccc}
 \hline\hline
 $\beta$ & $a^{-1}$ &
 $L^3\times T(\times L_s)$ & meas. &
 $am_{uds}$ & $am_c$ & $am_b$ \\
 \hline
 4.35 & 3.610(9) &
 $48^3\times 96(\times 8)$ & 400 &
 0.025 & 0.27287 & 0.66619 \\
 \hline\hline
 \end{tabular}
 \label{para}
 \caption{
 The parameters for our lattice calculation.  
 }
\end{table}
Our lattice setup is summarized in Table \ref{para}.
The lattice configurations are generated with $N_f = 2 +1$ flavors of quarks, which are formulated by the Mobius domain-wall fermion \cite{Brower:2012vk}.
The lattice spacing is $a^{-1} = 3.610(9)$~GeV, and each quark mass is set to $am_{uds} = 0.025$, $am_{c} = 0.27287$, and $am_b = 0.66619$.
This choice yields the meson masses $m_\pi = 714(1)$~MeV and $am_B = 3.44(1)$~GeV.
We insert two-different momenta ${\bf p}_{100} = (-\frac{2\pi}{L},0,0)$ and ${\bf p}_{110} = (-\frac{2\pi}{L},-\frac{2\pi}{L},0)$ for the final state of charmonium $c\overline{c}$.
For the initial $B$ meson state, we input a momentum ${\bf k} = (0,0,0)$.
The momenta ${\bf p}_{100}$ and ${\bf p}_{110}$ are smaller than the physical one, but we focus on these two inputs as a first step.
The energy spectrum with these input values are calculated as $E_K({\bf p}_{100})=855(3)$~MeV, $E_K({\bf p}_{110})=969(9)$~MeV, $E_{J/\psi}({\bf p}_{100})=3.127(1)$~GeV, and $E_{J/\psi}({\bf p}_{110})=3.158(1)$~GeV.
As we discussed in the definition of the $B \to K$ decay amplitude, the spectrum satisfies the condition $m_B<E_{J/\psi}+E_K$.
Namely, our setup does not suffer from artificial divergence, as we mentioned previously.
The source operators are set at $t_K = 42$ for K meson source, $t_J = 27$ for electromagnetic coupling $J_\mu$.
The B-meson source is set at the $t=0$.
In this study, statistical uncertainty is estimated with 100 independent configurations with four different source points per configuration.
Since the propagating mesons are heavy, the auto-correlation is not significant.

We use the renormalization constants determined through the moments of the corresponding three-point correlators \cite{Ishikawa_proc}.
They are $Z_{11} = 0.669(11)$ and $Z_{12} = 0.093(4)$.

\begin{figure}[tbp]
\begin{center}
 \includegraphics[width=5.2cm, angle=-90]{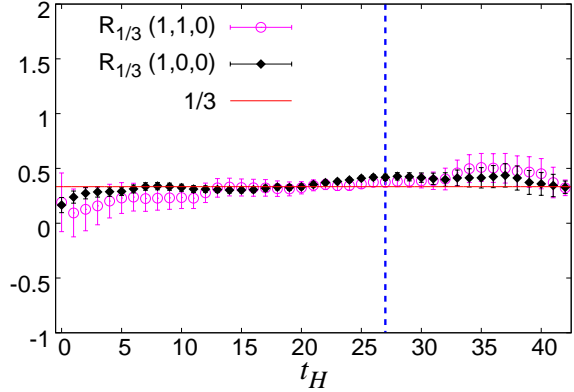}
 \caption{
 The ratio $R_{1/3}$ are shown for each input momenta. 
 The electromagnetic current is set at $t_J=27$ as shown by the dashed line.
}
\label{fig:R1/3}
\end{center}
\end{figure}

\begin{figure}[tbp]
\begin{center}
 \includegraphics[width=5.2cm, angle=-90]{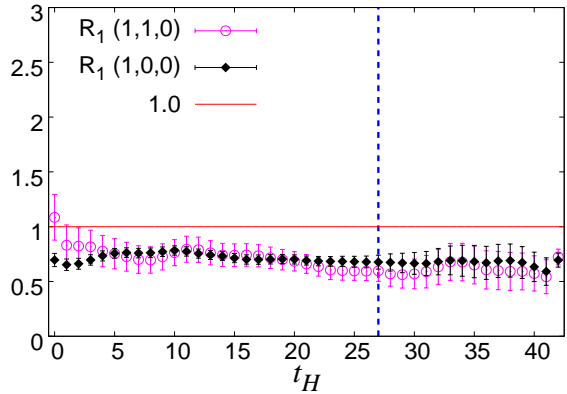}
 \caption{
 The ratio $R_1$ are shown for each input momenta. 
 The electromagnetic current is set at $t_J=27$ as shown by the dashed line.
}
\label{fig:R1}
\end{center}
\end{figure}

Figure \ref{fig:R1/3} shows the result for the ratio $R_{1/3}$, which should be equal to $1/3$ if the factorization is a good approximation.
The result is almost consistent with $1/3$, and we do not see any significant violation of the approximation.
On the other hand, the relation $R_1\simeq 1$ is not satisfied, as shown in Figure \ref{fig:R1}.
The size of the violation is as large as 30\%.
It suggests that the factorization approximation may underestimate the $\overline{c}c$ contribution to $B\to K\ell^+\ell^-$.

%There is a sizable deviation from the factorization approximation.
%Since this violation is significant for the theoretical prediction of the physical quantities, we have to take into account these nontrivial contributions, which are ignored in the factorization method.


\section{Discussions}

%We study the non-perturbative contribution to the $B\to K \ell^+\ell^-$ decay from the charmonium resonances.
%The factorization assumes these contributions small enough and then ignores them.
%We perform a lattice calculation to test this assumption, with a small b-quark mass.
%The renormalization constants are determined through the moments of three-point correlators.
%We calculate the four-point correlators and two-and three-point correlators to test the factorization on the lattice.
%There is a sizable deviation from the factorization approximation.

%Since this violation is significant for the theoretical prediction of the physical quantities, we have to take into account these nontrivial contributions, which are ignored in the factorization method.
%Since there are several studies of this non-factorizable contribution with phenomenological models, our lattice calculation could help us to understand the contribution, phenomenologically.

A quantitative estimate of the $c\overline{c}$ contribution to $B\to K^{(*)}\ell^+\ell^-$ remains a notoriously difficult task, because of the non-perturbative dynamics of QCD.
The first principle calculation of the lattice QCD can not be directly applied since there are many intermediate states that contribute to the real and imaginary parts of the amplitude.
In this work, we simplify the problem by considering an unphysical setup with a smaller $b$ quark mass, hoping that it captures the important part of the dynamics.
We find a significant violation of the factorization ansatz, which may be used as inputs for phenomenological models to study more realistic situations.
We also note that a large violation of factorization was previously found for the $K\to\pi\pi$ amplitude \cite{Boyle:2012ys}, which suggests the need for fully non-perturbative calculation for similar processes.

\section*{Acknowledgements}

The lattice QCD simulation has been performed on Blue Gene/Q supercomputer at the High Energy Accelerator Research Organization (KEK) under the Large Scale Simulation Program (Nos. 15/16-09, 16/17-14). 
Oakforest-PACS at JCAHPC under the support of the HPCI System Research Projects.
K. N. is supported by the Grant-in-Aid for JSPS (Japan Society for the Promotion of Science) Research Fellow (No. 18J11457).
This work is supported in part by the Grant-in-Aid of the Japanese Ministry of Education (No. 18H03710).

\bibliographystyle{JHEP}

\bibliography{Refs}

%\begin{thebibliography}{99}
%\bibitem{...}
%....

%\end{thebibliography}

\end{document}